# Machine Learning for Stock Prediction Based on Fundamental Analysis

Yuxuan Huang
*Broadridge Financial Solutions*
Toronto
Canada
yuxuan.huang@broadridge.com

Luiz Fernando Capretz
*Electrical and Computer Engineering*
Western University
Canada
lcapretz@uwo.ca

Danny Ho
*NFA Estimation Inc.*
Richmond Hill
Canada
danny@nfa-estimation.com

*Abstract* — **Application of machine learning for stock prediction is attracting a lot of attention in recent years. A large amount of research has been conducted in this area and multiple existing results have shown that machine learning methods could be successfully used toward stock predicting using stocks' historical data. Most of these existing approaches have focused on short term prediction using stocks' historical price and technical indicators. In this paper, we prepared 22 years' worth of stock quarterly financial data and investigated three machine learning algorithms: Feed-forward Neural Network (FNN), Random Forest (RF) and Adaptive Neural Fuzzy Inference System (ANFIS) for stock prediction based on fundamental analysis. In addition, we applied RF based feature selection and bootstrap aggregation in order to improve model performance and aggregate predictions from different models. Our results show that RF model achieves the best prediction results, and feature selection is able to improve test performance of FNN and ANFIS. Moreover, the aggregated model outperforms all baseline models as well as the benchmark DJIA index by an acceptable margin for the test period. Our findings demonstrate that machine learning models could be used to aid fundamental analysts with decision-making regarding stock investment.**

I. INTRODUCTION

The main motivation for predicting changes in stock price is the potential monetary returns. A large amount of research has been conducted in the field of stock performance prediction since the birth of this investment instrument, as investors naturally would like to invest in stocks which they have predicted will outperform the others in order to generate profit by selling them later. A large inventory of stock prediction techniques has been developed over the years, although the consistency of the actual prediction performance of most of these techniques is still debatable. The techniques for stock prediction can be classified into a small number of categories:

Fundamental analysis, where the predictions are made by studying the underlying companies through their published financial statements as well as macroeconomic factors.

Technical analysis, where the predictions are made by analyzing only the historical prices and volumes. Unlike fundamental analysts, who attempt to evaluate a stock's intrinsic value using publicly available information, technical analysts assume that a stock's price already reflects all publicly available information. There are three premises that technical analysis is based upon:

- Market action discounts everything
- Prices move in trends
- History repeats itself

Sentiment analysis, where the predictions are made by analyzing the published articles, reports and commentaries pertaining to certain stocks. Sentiment analysis is widely applied to different areas. For stock market, sentiment analysis is used to identify the overall attitude of investors towards a particular stock or the overall market.

Of the three general categories of stock prediction techniques, technical analysis and sentiment analysis are primarily used for short-term prediction on the scale of days or less. Fundamental analysis on the other hand, is used for mid-term and long-term prediction on the scale of quarters and years. In recent years, the popularity of applying various machine learning and data mining techniques to stock prediction has been growing. The majority of the existing studies using machine learning and data mining focus on creating prediction models based on technical analysis and sentiment analysis [1], [2], [3]. However, most of the short-term prediction models from many of the studies do not incorporate frictional cost in evaluation, the conclusiveness of the studies may be affected.

In this research, we aim to evaluate machine learning methods for long-term stock prediction based on fundamental analysis. We do so by comparing the prediction performance of three advanced machine learning methods based on fundamental analysis using fundamental features. To develop and test the machine learning models, we used data extracted from the quarterly financial reports of 70 stocks that appeared in the S&P 100 between 1996 and 2017. In order to evaluate the performance of different machine learning methods, we rank the

70 stocks based on their predicted relative return. Portfolios are constructed based on the ranking and the actual relative returns of the portfolios are used as the evaluating criteria.

## II. LITERATURE REVIEW

The majority of the existing studies that apply machine learning to stock prediction are based on technical analysis [2], [3]. Machine learning models developed in these studies take historical prices or technical indicators derived from historical prices as inputs. The popularity of technical analysis-based models is due to the popularity of technical analysis among the financial media and Wall Street financial advisers. In addition, stocks' technical data are available in much larger volume compared with financial fundamental data. This is because a stock's price and technical indicators are available with a daily sampling frequency, while its financial fundamental data is only published on a quarterly basis.

[4], [5], [6], [7], [8] explored different machine learning algorithms for short term stock price predication, including Artificial Neural Network (ANN), Support Vector Machine (SVM), Random Forest (RF), Neural Fuzzy Network and naïve-Bayes. For input data, historical price or technical indicators were used. These studies achieved various degrees of success.

A few studies on stock prediction and stock selection combined machine learning with fundamental analysis. Quah [9] compared three different machine learning models for stock selection based on fundamental analysis. The machine learning methods tested in this research are FNN, ANFIS and general growing and pruning radial basis function (GGAP-RBF). A dataset of 1630 stocks which were extracted within a period of ten years from 1995 to 2004 was used. Out of the ten years' annual data, only the last year's data were used for test set. Quah picked 11 of the most commonly used financial ratios as predictors based on Graham's book [10]. Instead of training the supervised learning models to do regression, Quah converted the prediction problem into a classification problem by classifying target variable into two classes. "Class 1" was defined as any stock which appreciates in share price equal to or more than 80% within one year, otherwise was classified as "Class 2". Such classification naturally creates an imbalanced dataset, as very few stocks are able to appreciate over 80% within one year in any given year. Therefore, an over-sampling technique was used on the minority class in order to balance the dataset. Over-sampling was used on training set only for the purpose of avoiding data scooping. According to the experimental results, both FNN and ANFIS models were able to achieve above market average annual appreciation of selected stocks at 13% and 14.9% respectively. The average annual appreciation of the market for the test set is 11.2%. On the other hand, GGAP-RBF performed poorly. The author also mentioned in the conclusion that the availability of financial data is a major limitation of this study.

A recent study by Namdari and Li [11] also used FNN on stock trend prediction. They used 12 selected financial ratios of 578 technology companies on Nasdaq from 2012-06 to 2017-02 as their dataset. Instead of simply normalizing or standardizing these continuous features, they discretized all features by conducting topology optimization. For comparison, they also developed a different FNN model for predicting stock price trend based solely on historical price for the same companies and the same period of time. The results suggest that the FNN model based on fundamental analysis was able to outperform the alternative model based on technical analysis with overall directional accuracy of 64.38% and 62.84%, respectively.

Bohn [12] combined technical analysis, fundamental analysis and sentiment analysis and compared a set of machine learning models for long-term stock prediction. He used a universe of around 1500 stocks which appear in the S&P 500 between 2002 and 2016 for experiment. Regression models were built, and ranks were induced based on the model predictions for each validation and test week. He evaluated the model performance using the Spearman rank correlation coefficient between predicted rank and actual rank. The results suggest that the neural network model combined with iterative feature selection could match the performance of a model developed with human expertise from an investment firm.

Yu et al. [13] developed a novel sigmoid-based mixed discrete-continuous differential evolution algorithm for stock performance prediction and ranking using stock's technical and fundamental data. The evaluation metrics and feature selection process used in this study is the same as in [12]. 483 stocks listed in Shanghai A share market from Q1 2005 to Q4 2012 were used for model building and testing. The results suggest that the proposed model can create portfolios that significantly outperform the benchmark.

Previously, we had experimented with FNN and ANFIS for Function Point calibration [14]. The results were encouraging so that we would like to expend upon that in this work.

## III. BACKGROUND

### A. Feed-forward Neural Network

Feed-forward Neural Network (FNN), or Multi-Layer Perceptron (MLP), is the simplest and very versatile form of neural network architecture. An FNN consists of at least three layers: an input layer, a hidden layer and an output layer. The supervised learning technique of gradient descent is used for backpropagation. There are many hyperparameters that can be tuned during the model validation of an FNN in order to achieve the optimal model generalization, including weight initialization method, learning rate, number of hidden layers, number of hidden unites in each hidden layer, activation function, etc.

### B. Random Forest

Random Forest (RF) is a flexible supervised learning algorithm which can be used for both classification and regression tasks. It builds multiple decision trees during the data fitting process. For generating results, RF takes the mean value of the output of all decision trees for a regression problem. For classification problems, the majority voting from the decision trees is used as the result. Many hyperparameters can be tuned to increase the performance of RF, including number of estimators, minimum sample split, maximum features, etc.

### C. Adaptive Neural Fuzzy Inference System

ANFIS is an instance of the more generic form of the Takagi-Sugeno-Kang (TSK) fuzzy inference system. It replaces the fuzzy sets in the implication with a first order polynomial equation of the input variables [15]. The ANFIS system consists of rules in IF-THEN form. In general, there are five different layers in an ANFIS system. Layer 1 converts each input value to the outputs of its membership functions:

$$O_i = \mu_{A_i}(x) \quad (1)$$

where is the input to node and is the bell-shaped membership function with maximum equal to 1 and minimum equal to 0.

Layer 2 calculates the firing strength of a rule by simply multiplying the incoming signals.

Layer 3 normalizes the firing strengths:

$$\overline{w_i} = \frac{w_i}{\sum_j w_j} \quad (2)$$

Layer 4 consists of adaptive nodes with function defined as:

$$O_i = \overline{w_i}(p_i x + q_i y + r_i) \quad (3)$$

where $\overline{w_i}$ is the normalized firing strength from the previous layer and $(p_i x + q_i y + r_i)$ is the first order polynomial with three consequent parameters $\{p_i, q_i, r_i\}$.

Layer 5 takes the weighted average of all incoming signals and delivers a final output:

$$O_i = \sum_i \overline{w_i} f_i = \frac{\sum_i w_i f_i}{\sum_i w_i} \quad (4)$$

where $f_i$ if the first order polynomial mentioned above.

Tuning an ANFIS involves determining the number of membership functions for each input and the type of input membership function.

IV. METHODOLOGY

A. Data Preparation

Sample stocks used for this experiment were chosen from the S&P 100 Index components. The index includes 102 leading U.S. stocks which represent about 51% of the market capitalization of the entire U.S. equity market. Because the composition of the S&P 100 index is frequently revisited, we decided to use its components as of December 2018 [16]. Historical financial data for each of the S&P 100 components were retrieved online in csv format [17]. These data were extracted from companies' SEC 10_Q filings, which are published quarterly. The original dataset has large blocks of missing values concentrated on a few features, while other missing values were sparsely populated across the entire dataset. We eventually decided to use a combination of feature deletion and mean substitution. In cases where a fundamental factor had large blocks of missing values or over 50% values missing, it was removed. We also removed some non-fundamental features such as price high and price low. After the feature dropping, there were still some sparsely located missing values which account for less than 3% of total samples. These missing values were then substituted by the average of the two adjacent values. For example, if the revenue data for 2015Q3 is missing, it is substituted by the mean of the revenue values of the 2015Q2 and 2015Q4.

Our target variable in this research is stock's quarterly relative returns with respect to the Dow Jones Industrial Average (DJIA). The major benefit of using relative return versus simple absolute return is that we are able to filter out some factors affecting the broader market by subtracting overall market performance from the performance of an individual stock. Many features from the raw dataset possess a clear global trend with respect to time. As we transfer this time series problem into a supervised learning problem, these features with global trends could hinder our machine learning models' ability to generalize and provide reliable predictions. We therefore took the percentage change between consecutive observations for all features, which is calculated as follows:

$$\Delta x_t = \frac{(x_t - x_{t-1})}{x_{t-1}} \times 100\% \quad (5)$$

After removing the trends from input variables, we performed dataset partition and standardization. The dataset was partitioned into train/validation/test in the proportion of 60%/20%/20%. Then we first standardized the train set following Equation (6). The validation set and test set were standardized using the mean and standard deviation from the train set in order to prevent data snooping. From a time series perspective, data from Q1 1995 to Q1 2008 was used for training; data from Q2 2008 to Q2 2013 was used for validation, and data from Q3 2013 to Q4 2017 was used for testing. Moreover, we trained the models with the training set and the validation set combined after model validation for generating the final test results on the test set. This helps us to maximize the usage of data for training the models.

$$x' = \frac{x - \bar{x}}{\sigma} \quad (6)$$

After the data preparation process was completed, we ended up with 21 features and 70 stocks. Each stock has 88 observations, ranging from Q1 1996 to Q4 2017, with an interval of one quarter between two consecutive observations. The 21 features are illustrated in Table 8.

B. Local Learning

We tried both building a single model for all stocks and building one model for each stock. The two approaches can be classified as global learning and local learning. Models trained with global learning enjoy a larger set of training data, while models trained with local learning are more task specific and usually enjoy better performance [18]. Local learning approach was proven to have better performance in our early experiment, and thus we built one model for each stock for all three algorithms.

C. Evaluation Metrics

The goal of this project is to develop a system which can be used to guide stock portfolio design strategy for long term investment. Therefore, simple and general evaluation methods are preferred. We decided to build regression models to predict the price for each stock, and then induce a ranking of the stocks by sorting their predicted relative returns. The ranking can then be used for portfolio design, and the actual performance of the portfolios in terms of real relative return can be evaluated with ease.

When training a regression model, the metric or the loss function depends on the specific algorithm. Moreover, the loss function used in model training is also a hyperparameter which can be tuned. For the FNN and ANFIS models, we use RMSE as the training loss function. The RF algorithm, unlike FNN and ANFIS, does not involve training cycles and loss function.

After fitting a model with the training data, it is then evaluated on the validation data. The stocks are ranked by their predicted relative returns for each of the quarters. The top one third stocks with the highest ranking are selected into a portfolio. The real relative return of the selected portfolio for each quarter is then calculated, assuming the portfolio is equal weight. The average real relative return of the equal weight portfolio is calculated with the formula:

$$\overline{R_p} = \frac{1}{\#quarters}\sum_{q=1}^{\#quarters} R_p(q) \quad (7)$$

where $R_p(q)$ is the real relative return of the selected equal weight portfolio for quarter calculated as follows:

$$R_p(q) = \frac{1}{\#stocks}\sum_{i=1}^{\#stocks} R_i(q) \quad (8)$$

where $R_i(q)$ is the real relative return of a stock in the selected portfolio for the single quarter $q$.

If the performance of the portfolio selected by our model is highly volatile from quarter to quarter, even if it can produce good relative return on average, it might still be undesirable. This is because high volatility leads to high risk, and high volatility can also diminish compounding return in long term. In the financial world, the Sharpe ratio is commonly used to help investors understand the return of an investment compared to its risk. The Sharpe ratio is a risk adjusted return ratio calculated as follows:

$$\text{Sharpe Ratio} = \frac{R_p - R_f}{\sigma_p} \quad (9)$$

where $R_p$ is the return of the portfolio, $R_f$ is the risk-free rate, and $\sigma_p$ is the standard deviation of portfolio return over the considered duration. When constructing a portfolio, investors want to maximize the Sharpe ratio of the portfolio in order to get the maximum return with the minimum risk. For this project, we use a modified version of the Sharpe ratio as our risk-adjusted relative return metric:

$$\text{Portfolio Score} = \frac{\overline{R_p}}{\sigma_p} \quad (10)$$

where $\overline{R_p}$ is calculated as in Equation 7. The risk-free rate is left out for simplicity.

*D. Feature Selection*

In this project, the RF algorithm is used for feature selection. The RF algorithm has demonstrated its efficiency in feature selection from previous studies [19] [20]. The algorithm is applied on the training data of all stocks in order to obtain estimates of feature importance of each feature. The most important features are then selected for model building.

*E. Model Agreegation*

After each individual algorithm is tested and evaluated, the bootstrap aggregating algorithm is applied in order to assemble the prediction results of different algorithms with the goal of improving stability and accuracy. Bootstrap aggregation is a simple and widely used meta-algorithm for aggregating predictive models. It is also the algorithm used in Random Forest for aggregating results from individual decision trees into a final output.

In this project, bootstrap aggregation is used on the rankings of stocks produced by each algorithm. A stock is selected for inclusion in the portfolio if the majority of the algorithms predict that the stock's performance for the next quarter is in the top one third of all stocks.

V. RESULTS AND DISCUSSION

*A. Baseline Models*

For this phase of the experiment, the machine learning algorithm: FNN, ANFIS and RF are trained to predict the quarterly relative return of each of the 70 stocks. A rank of the stock is then induced from the predicted relative returns for each quarter. The ranking is then used for portfolio building. Before being used to produce predictions on the test set, each algorithm is first validated on the validation set for hyperparameter tuning.

For evaluation, portfolios consisting of stocks from the top and bottom of the ranking are both evaluated. The reason for including the portfolios consisting of stocks with the worst predicted performance in our evaluation is that if our models can successfully identify the worst performing stocks, profit could potentially be generated by shorting these stocks.

We compare the experimental results for different machine learning algorithms. The "Top20" and the "Bottom20" portfolios are used for cross-model comparison, because they represent roughly the top one third and the bottom one third of the universe.

Moreover, the compounded relative return over the test period of 18 quarters are calculated for each method as an additional metric. The results are succinctly presented in Table 1 and Table 2.

TABLE I
BASELINE MODEL RESULTS FOR "TOP20 BUY"
PORTFOLIOS

|        | Mean     | STD  | Portfolio Score | Compound |
|--------|----------|------|-----------------|----------|
| FNN    | 0.831%   | 4.11 | 0.202           | 14.4%    |
| ANFIS  | 0.621%   | 5.59 | 0.111           | 8.85%    |
| RF     | 1.63%    | 3.93 | 0.414           | 32.1%    |
| Universe | -0.0164% | 3.55 | -0.00460        | -1.35%   |

TABLE II
BASELINE MODEL RESULTS FOR "BOTTOM20 SELL"
PORTFOLIOS

|        | Mean     | STD  | Portfolio Score | Compound |
|--------|----------|------|-----------------|----------|
| FNN    | -0.768%  | 4.22 | -0.182          | -14.3%   |
| ANFIS  | -0.506%  | 4.29 | -0.118          | -10.2%   |
| RF     | -1.39%   | 4.57 | -0.305          | -23.7%   |
| Universe | -0.0164% | 3.55 | -0.00460        | -1.35%   |

The observations based on the experimental results are as follows:

1) All "Buy" portfolios outperform the universe in terms of average quarterly relative return, Portfolio Score and compound relative return by a significant margin. On the other hand, all "Sell" portfolios underperform the universe in terms of the same metrics. Therefore, we can posit a safe conclusion that all three supervised learning models are able to predict, with a good degree of accuracy, the near-term winners and losers from a universe of stocks based on the stocks' most recent fundamental financial ratios. The results obtained challenge both the weak and the semi-strong form of the well-known EMH (Efficient Market Hypothesis).

2) The RF outperforms other models in constructing both "Buy" and "Sell" portfolios in terms of all evaluation metrics by a significant margin, with the exception of standard deviation of its "Sell" portfolio. The "Buy" portfolio constructed by RF achieved a mean quarterly relative return of 1.63%, compared with the mean quarterly relative return of the universe at -0.0164%. The compound relative return outperforms the universe by 33.5% over the test period of 18 quarters or four and half years.

3) The ANFIS underperforms other models. The result could be because of the huge number of parameters to be tuned during the training process of the ANFIS model and the limited volume of training data. For instance, for a fuzzy inference system with 10 inputs, each with two membership functions, the ANFIS could generate 1024 (=2^10) rules. In our case, we have 21 inputs. More training data or fewer input features could potentially improve the prediction performance of ANFIS.

4) The standard deviations of quarterly relative returns for the selected portfolios are higher than that of the universe. This means the selected portfolios are more volatile than the benchmark. This is expected as the smaller number of stocks in a portfolio naturally leads to higher volatility.

5) All models seem to be better at identifying winner than identifying losers by a small margin. More investigation is required to find the reasons behind such a phenomenon.

*B. Applying Feature Selection*

The RF regressor is applied on the test data of all stocks for feature selection. The feature importance, as well as its standard deviation, are calculated for each feature. The features are then ranked according to their feature importance, as illustrated in Figure 1. The red bars represent feature importance and the black lines represent the standard deviation.

The primary reason for applying feature selection is to reduce model complexity of FNN and ANFIS models and mitigate potential overfitting. Using only the top 2 most important features could reduce model complexity significantly. However, we would also face the consequence of significant loss of information if we drop all other features. We decide to use the top 6 most important features for experiment as a balance between model complexity and information loss. The 6 selected features are illustrated in Table 3.

The FNN, ANFIS and RF models are validated and tested following the same procedure with the selected features. The test results are presented in Table 4 and Table 5.

The observations based on the experimental results are as follows:

1) Feature selection improves the prediction performance of FNN significantly. The Portfolio Score of the "Buy" portfolio improves from 0.202 to 0.274, and the Portfolio Score of the "Sell" portfolio improves from -0.182 to -0.302.

2) The Portfolio Score of the "Buy" portfolio produced by ANFIS improves from 0.111 to 0.159 with feature selection, while the "Sell" portfolio does not see an improvement.

3) Feature selection does not help to improve the performance of the RF model. In fact, the performance of RF is worsened with selected features.

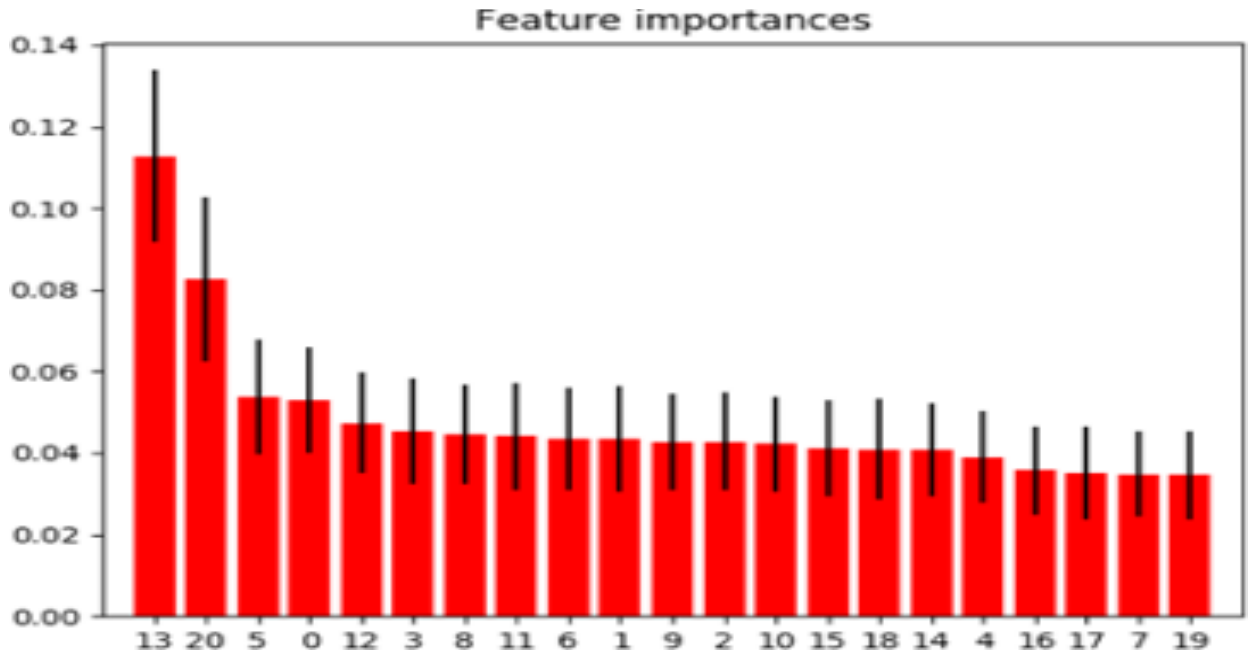
Fig. 1. Feature importance based on RF

TABLE III. TOP SIX FEATURES SELECTED BY RF

| No. | Feature Name |
|---|---|
| 13 | PB |
| 20 | Relative Return |
| 5 | Book Value |
| 0 | PE |
| 12 | Capital Expenditure |
| 3 | Liability |

TABLE IV
RESULTS FOR "TOP20 BUY" PORTFOLIO WITH
SELECTED FEATURES

|  | Mean | STD | Portfolio Score | Compound |
|---|---|---|---|---|
| FNN+FS | 0.990% | 3.62 | 0.273 | 18.1% |
| ANFIS+FS | 0.647% | 4.06 | 0.159 | 10.8% |
| RF+FS | 1.14% | 3.66 | 0.310 | 21.2% |
| Universe | -0.0164% | 3.55 | -0.00460 | -1.35% |

TABLE V
RESULTS FOR "BOTTOM20 SELL" PORTFOLIO WITH
SELECTED FEATURES

| | Mean | STD | Portfolio Score | Compound |
|---|---|---|---|---|
| FNN+FS | -1.45% | 4.80 | -0.302 | -24.6% |
| ANFIS+FS | -0.232% | 3.99 | -0.0582 | -5.41% |
| RF+FS | -0.877% | 4.32 | -0.203 | -16.0% |
| Universe | -0.0164% | 3.55 | -0.00460 | -1.35% |

*C. Model Aggregation*

In order to further improve prediction accuracy and stability, the predictions of the best performing models are aggregated using bootstrap aggregation. We decided to use the FNN and ANFIS models with feature selection and RF model without feature selection for final aggregation, because of the fact that feature selection does not improve the performance of RF. We tested two aggregation strategies: "agg2" and "agg3". In "agg2", for a stock to be selected into the "Buy" portfolio, there had to be at least 2 out of the 3 models that ranked the stock in the "Top20" for the quarter. In "agg3", all 3 models had to rank a stock in the "Top20" in order for the stock to be selected. Such aggregation naturally leads to fewer stocks being selected into the "Buy" portfolio for each quarter. For example, there could only be 10 stocks, which are in "Top20" portfolios for all three models, in the "Buy" portfolio for "agg3" in a test quarter. The smaller number of stocks in the aggregated portfolios could lead to higher volatility or standard deviation of performances over the test period. The relative returns of "agg2" and "agg3" for every test quarter are illustrated in Figure 2 and Figure 3. We can clearly see that "agg3" performs much better than the universe as well as "agg2" from the figures.

As we can see from the results presented in Table 6 and Table 7, "agg3" outperforms all individual models, as well as "agg2", for constructing both "Buy" and "Sell" portfolios. The "Buy" portfolio constructed by "agg3" achieves a mean quarterly relative return of 5.11%, a Portfolio Score of 0.759 and an impressive compounded relative return of 137% over the course of 18 test quarters.

VI. CONCLUSION

In this study, we looked at the problem of predicting stock performance with machine learning methods. Although a substantial amount of research exists on this topic, very few aims to predict stocks' long-term performance based on fundamental analysis. We prepared 22 years' worth of stock financial data and experimented with three different machine learning methods for long term stock performance prediction. In addition, we applied feature selection and bootstrap aggregation in order to improve the prediction performance and stability.

To produce effective and reliable models, we faced two major challenges. The first challenge was to put together a sizable dataset for experimenting. Due to the fact that publicly traded companies only publish their financial data on a quarterly basis and the relatively short history of digitally archiving these data, we did not have as much data as we wanted to work with. We extracted as much data as we could for 70 large-cap stocks which are S&P 100 components. The original dataset consisted of a large number of missing values, and we went through a series of data preprocessing steps in order to prepare the data for model training and testing. We experimented with building one model for all stock and building one model for each stock, and we decided on using the second approach for all algorithms based on early experimental results. The second challenge involves market efficiency, which places a theoretical limit on how historical patterns in the stock market could be used for predicting its future behavior. We took several measures to deal with this challenge. Firstly, we carefully split our data into training, validation and testing sets and made sure that we did not accidentally snoop the test data or overfit the models. Secondly, we used the Portfolio Score as our primary validation and evaluation metric. The Portfolio Score takes into account not only the performance of the constructed portfolio, but also its standard deviation over the validation period. Finally, we also employed the feature selection technique in order to remove unreliable features and reduce model complexity.

The experimental results we presented show that all three machine learning methods we experimented with are capable of constructing stock portfolios which outperform the market without any input of expert knowledge, if fed with enough data. Out of the three algorithms, RF achieves the best performance.

### TABLE VI
### RESULTS FOR "BUY" PORTFOLIOS

|         | Mean     | STD  | Portfolio Score | Compound |
|---------|----------|------|-----------------|----------|
| FNN+FS  | 0.990%   | 3.62 | 0.274           | 18.1%    |
| ANFIS+FS| 0.647%   | 4.06 | 0.159           | 10.8%    |
| RF      | 1.63%    | 3.93 | 0.414           | 32.1%    |
| Agg2    | 1.45%    | 3.85 | 0.385           | 28.0%    |
| Agg3    | 5.11%    | 6.73 | 0.759           | 137%     |
| Universe| -0.0164% | 3.55 | -0.00460        | -1.35%   |

### TABLE VII
### RESULTS FOR "SELL" PORTFOLIOS

|         | Mean     | STD  | Portfolio Score | Compound |
|---------|----------|------|-----------------|----------|
| FNN+FS  | -1.45%   | 4.80 | -0.301          | -24.6%   |
| ANFIS+FS| -0.232%  | 3.99 | -0.0582         | -5.41%   |
| RF      | -1.392%  | 4.57 | -0.305          | -23.7%   |
| Agg2    | -1.77%   | 4.89 | -0.362          | -29.0%   |
| Agg3    | -2.32%   | 6.93 | -0.335          | -37.2%   |
| Universe| -0.0164% | 3.55 | -0.00460        | -1.35%   |

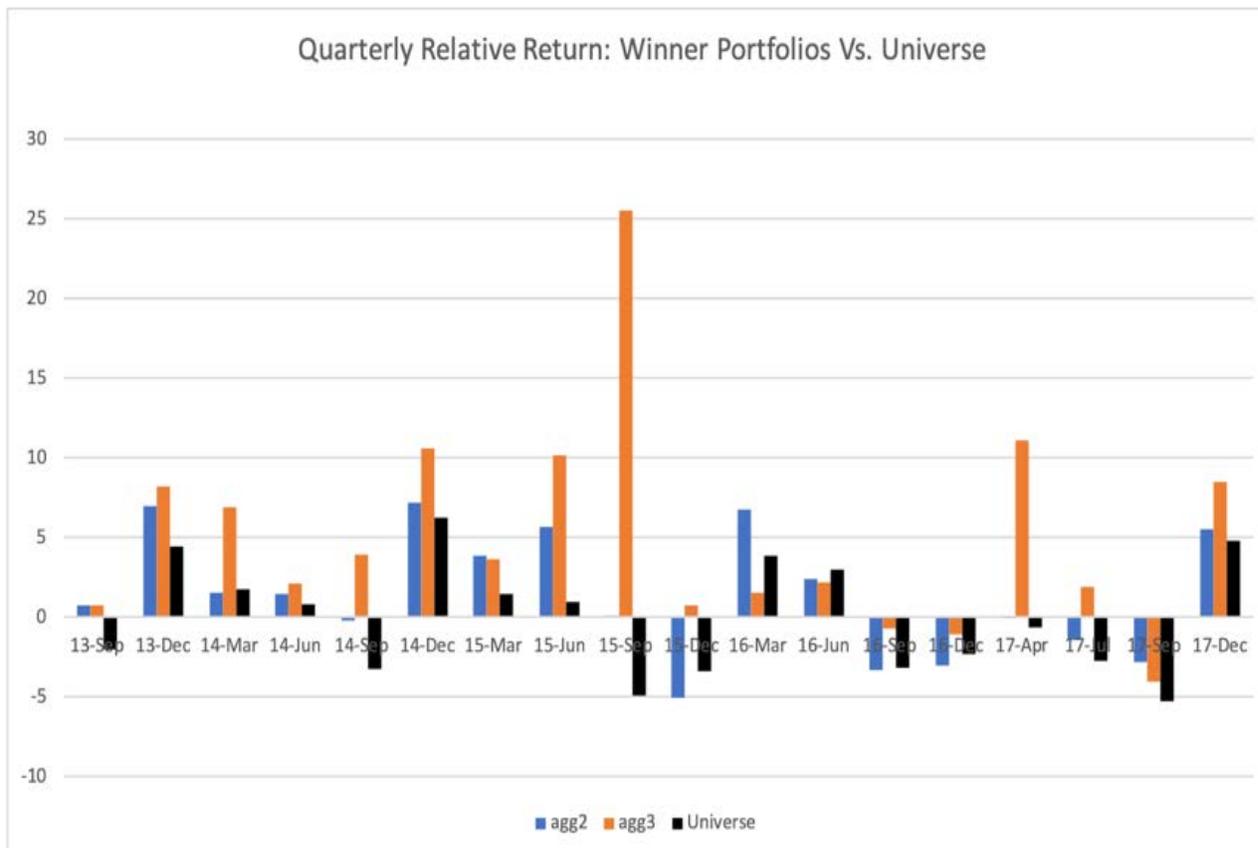

Fig. 2. Relative return of aggregated "Buy" portfolio

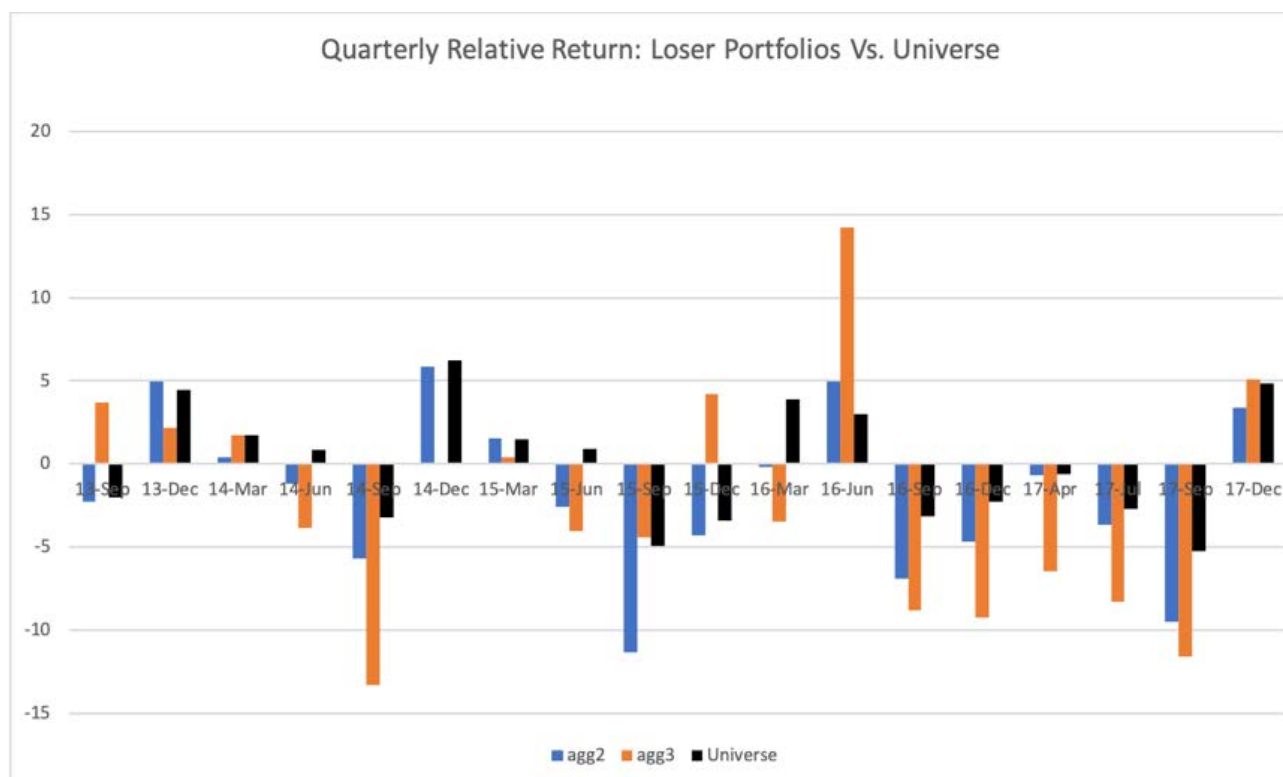

Fig. 3. Relative return of aggregated "Sell" portfolio

By applying feature selection and aggregating the different algorithms, our aggregated model achieves a Portfolio Score of 0.759 and -0.335 for the "Buy" and "Sell" portfolios respectively. This shows that our model could help to build portfolios which outperform the benchmark using historical financial data.

There are many limitations to this research, and there are many ways this topic could be further explored. First of all, the prediction performances of the models used in this research are largely restricted by the limited volume of available data. More data could potentially improve model performance as well as conclusiveness of our results. We used simple standard deviation of portfolio returns as a factor to measure risk. More rigorous stock covariance matrix analysis could be applied. More algorithms could be tested, such as different variations of neural network. Validation mechanisms such as cross validation, sliding window and expanding window could be applied to improve the model validation process and potentially improve model generalizability and robustness. Moreover, different feature selection methods could be explored, such as iterative feature selection. The number of features selected after feature selection is rather arbitrary. We could experiment with different number of most important features to further improve the effectiveness of our feature selection. We could also try to incorporate technical analysis and sentiment analysis in our model.

TABLE VIII
INPUT VARIABLES AFTER DATA PREPARATION

| No. | Feature Name | |
|---|---|---|
| 0 | $PE$ | % change of price per earning |
| 1 | $Assets$ | % change of total assets |
| 2 | $Current\_assets$ | % change of current assets |
| 3 | $Liabilities$ | % change of total liabilities |
| 4 | $Current\_liabilities$ | % change of current liabilities |
| 5 | $Book\_value$ | % change of book value |
| 6 | $Revenue$ | % change of revenue |
| 7 | $Earning$ | % change of earning |
| 8 | $Cash\_from\_Op$ | % change of cash from operation |
| 9 | $Cash\_from\_Inv$ | % change of cash from investment |
| 10 | $Cash\_from\_fin$ | % change of cash from financing |
| 11 | $Cash$ | % change of cash |
| 12 | $Capital\_Exp$ | % change of capital expenditure |
| 13 | $PB$ | % change of price per book |
| 14 | $Cash\_per\_share$ | % change of cash per share |
| 15 | $Current\_ratio$ | % change of current ratio |
| 16 | $Net\_margin$ | % change of net margin |
| 17 | $ROA$ | % change of return on assets |
| 18 | $Asset\_turnover$ | % change of asset turnover |
| 19 | $EPS$ | % change of earning per share |
| 20 | $Relative\_return$ | Past quarter relative return on price |